# THE ROLE OF SOCIAL MEDIA ON SELECTED BUSINESSES IN NIGERIA IN THE ERA OF COVID-19 PANDEMIC

**Cajetan Ihemebiri[1], Elochukwu Ukwandu[1], Lizzy Ofusori[2] and Comfort Olebara[3]**

[1]Department of Applied Computing and Engineering, Cardiff School of Technologies, Cardiff Metropolitan University, Cardiff, Wales, United Kingdom

[2]School of Management, Information Technology & Governance,
University of KwaZulu-Natal, Westville Campus, Durban, South Africa.

[3]Department of Computer Science
Faculty of Physical Science, Imo State University, Owerri, Imo State, Nigeria.

Correspondent Author: Elochukwu Ukwandu, eaukwandu@cardiffmet.ac.uk

**Abstract**
As several countries were experiencing unprecedented economic slowdowns due to the outbreak of COVID-19 pandemic in early 2020, small business enterprises started adapting to digital technologies for business transactions. However, in Africa, particularly Nigeria, COVID-19 pandemic resulted to some financial crisis that impacted negatively on the sustainability of small and medium-sized (SMEs) businesses. Thus, this study examined the role of social media on selected SMEs in Nigeria in the heat of the COVID-19 pandemic that led to several lock downs in a bid to curtail the spread of the virus. Cross-sectional survey research design was used alongside convenience population sampling techniques. The population was categorised based on selected SMEs businesses, while a quantitative research approach was adopted, and primary data were collected using a questionnaire. The questionnaires were administered to owners and operators of SMEs in Ikotun and Ikeja areas of Lagos State, Nigeria. A total of 190 questionnaires were distributed, where 183 usable responses were analysed. The findings of the study show that SMEs were aware of the usefulness of social media to their businesses as they largely leveraged it in conducting their businesses during the national lockdowns. The study recommended that labour/trade unions should sensitise and encourage business owners on the benefits of continuous use of social media in carrying out their business transactions.

## 1. Introduction/background

Coronavirus (COVID-19) identified in China around late 2019 became pandemic in 2020 and the unfavourable situation created by the pandemic had significant impacts on the entire world (Adenomon and Maijamaa, 2020). On February 27, 2020, the first COVID-19 case was confirmed in Lagos State, Nigeria. In line with this discovery, the federal government of Nigeria issued directives aimed at prohibiting mobility in Lagos State until the end of March 2020. These resulted to limitations on entry and exit as well as mobility within the states in Nigeria. As the economy came to a halt, worst-case scenarios became a reality, because of containment efforts and spill-over effects from the actual economy to financial markets, investors were absorbing the consequences of interrupted supply chains (Gill & Johnson, 2010). As a result, businesses were turning to technologies and online marketing to solve the problem of in-person shopping (Oji, 2020). Majority of past research on the impact of COVID-19 pandemic concentrated on preventative health, but just a few have looked at the impact of the pandemic on wellbeing and continuity of enterprises such as schools, shopping malls, and restaurants. This represents a gap in the literature and gives an opportunity for this study to address. It should be noted that various business ventures around the world especially SMEs use social media outlets in conducting their entrepreneurial activities. This has led to business start-ups having access to customers outside their local area, thus uncovering new markets, unlimited flow of information and ideas. Social media enables real-time sharing, promotion, and discussion of products and services across social network channels to a large audience (Spears, 2007). During COVID-19, it was observed that numerous Nigerian firms were unaware of the capabilities of social media in their operations. Hence, this study aims to assess the role of social media on the selected SME businesses in Nigeria in the era of the COVID-19 pandemic by answering the following research questions:

1. At what level were SMEs in Nigeria aware of the usefulness of social media to their businesses during the COVID-19 pandemic national lockdowns?
2. At what extent did SMEs in Nigeria leverage the potentials of social media on their businesses during the pandemic lockdowns?
3. To what degree did social media usage impact businesses during the COVID-19 pandemic induced lockdowns?

## 2. Literature Review
### 2.1 Social media usage for businesses during the COVID-19 pandemic

In the wake of COVID-19 pandemic businesses were seen developing variety of measures to keep their sales and transaction afloat due to social distancing measures used to curb the spread of the virus. Various organisations around the world were seen using social media as vital tool to ensure hitch-free business operations and according to Yan (2018), 80% of these organisations used various social media platforms. The result also showed that 75% of these used LinkedIn for recruitment, while using Facebook, Twitter, Facebook, and Instagram primarily for marketing and advertisement. According to Margado (2020), during the pandemic Slovakia saw a significant increase in e-commerce turnover due to the potential of social media on SMEs as between March 2019 to March 2020, online sales climbed by 44%. For instance, e-shops selling medical products such as cleaning agents and facial masks saw a 130% rise in orders year over year during the first lockdown.

Svatosova (2015) findings showed that SMEs in United States of America (USA) largely use social media platforms in advertising their goods and services and as a medium to communicate with customers, whereas organisations in developing countries such as the Czech Republic neglect social media marketing, resulting in their lack of success in the global market. In Nigeria, Ile *et al*. (2018) evaluated the use of social media as a platform for effective functioning of small and medium-sized firms (SMEs) in Nigeria in the twenty-first century. From their study, SMEs based in Anambra State, Nigeria ranked highest in the innovative use of social media for customer engagement. However, due to COVID-19 pandemic, the financial crisis had impact on the sustainability of SMEs businesses. The pandemic prompted the government to advise citizens to avoid being present physically and instead choose for online shopping. Unfortunately, many Nigerian SMEs were still unfamiliar with internet marketing (Sugandini *et al.,* 2019). These SMEs had to close their doors due to a drop in sales; however, this can be avoided by using online platforms for advertising, sales, and purchases of commodities (Oji, 2020). As a result, the research sheds light on the impact of social media on selected SMEs in Nigeria during the COVID-19 pandemic.

### 2.2 Impact of social media on businesses – Charles Okoro
According to Manoj (2017), Social media utilisation has various impacts on businesses and the opportunities it presents which include: an increase in revenue, improvement of brand/ awareness creation, networking, and recruitment. In addition, Susanto *et al*., (2020) reveals six positive impacts of social media on businesses. First, social media provides improved communication between business administrators and clients - customer service. Secondly, customers can be easily attracted and be recommended through social media. Thirdly, it is a cheap mode of advertising as well as promoting goods and services. Fourthly, it helps build a favoured image for an organisation. Fifthly, the latest trends are displayed in social media to help users advance their strategies. Finally, profit generated from traffic on social media platforms can be channelled back into the business.

Timilsina (2017) observed that social media has a positive influence on restaurant business in their study to investigate the impact of social media usage in business and how this is affecting businesses using Oulu-based restaurants for the survey. Restaurants that use these digital channels see a rise in sales and increased client awareness and Instagram and Facebook are social media instruments used mostly by restaurants. Also, microbusinesses use social media because it allows them to reach out to individuals easily and quickly, thus allowing them to build relationships and contact potential clients. Most people use more than two social media platforms to learn about different brands. However, despite its benefits, during the pandemic transporting goods to customers was a difficult task, as obstacles like lockdown can impede dispatch riders from arriving on time and even block deliveries (United Nations Conference on Trade and Development, 2020). Also, the greater number of social media users are the younger generation, ranging ages 15 to 34, and their reactions to some products may be discouraging. Another shortcoming of social media is false information about competitors' products, and this can be distributed through social media, depending on data from consumers' profiles (Svatosova, 2015).

## 2.3 Research theories

This study adopted the diffusion of innovation theory and gratification theory. The diffusion of innovation describes how content or ideas expand across time in life through various channels and social systems (Katz *et al*, 1963 cited in Ojobor, 2002). The theory states that for a new idea to spread, it must go through stages of awareness, interest, appraisal, trial, and acceptance. The information diffusion model included four stages: information, persuasion, choice or adoption, and confirmation (McQuail, 2011). The importance of this theory to this project arises from the fact that there is an increase in social media usage, and as a result, various users may embrace them in diverse ways. While the usage and gratification hypothesis discuss how various people utilise media and what pleasures they get from it. According to Folarin (2005), the audience of media messages is seen as proactively impacting the consequence phase and absorbs media content based on his wants and views. This means that individuals of the public may proactively choose and employ media information to meet their requirements and gratify their interests and purposes. It is assumed that persons who choose to patronise goods on social networks use it in the hopes of receiving pleasure from those goods. All those who don't buy the goods may have not found the messaging to be satisfying. Hence gratification theory is important to this study as businesses also utilise social media in this era of COVID-19 pandemic.

## 3. Research methodology

The study used a cross-sectional research design as its research method, the objectives and the set of research questions influenced this decision. Inferential statistics were used to make possible conclusions and decisions, whereas descriptive statistics were employed to display the facts in a digestible fashion. The design allowed for the comparison of outcomes with the purpose of improving the dependability of the study's conclusions. The study sites for this study are Ikotun and Ikeja in Lagos, Nigeria. The two locations were chosen because of their contribution to the state budget and home to most SMEs in Lagos. The study populations are Small Medium-sized business owners and technicians in Ikotun & Ikeja, Lagos State, Nigeria. The sample size was determined via Convenience sampling technique as described in Etikan *et al.* (2016), Stratton (2021). Convenience samples is a non-probabilistic sampling technique whose elements may be selected accidentally, spatially, or administratively. As this sampling technique is less objective than probability techniques, the researchers' chose it because of the prevailing social distancing circumstances during the lockdowns as getting the target population to participate were dicey, so the participants were either approached remotely by the researcher, referred, or self-selected to participate in the study. In all the researchers made efforts to have a heterogeneous representation of the target SMEs in the research locations.

## 3.1 Data collection techniques

A questionnaire was used for the primary data collection. The outcome of the study was analysed with a statistical analysis SPSS tool that examines the relationship of variables that are considered in this research work. There was a total of twenty-six questions in the survey (26). The questions were divided into four (4) sections, the first of which was aimed to capture the participants' socio-demographic data, including personal information like sex, age bracket, level of education, and so on. The second section uses a dichotomous option to determine the level of awareness of the SMEs about the value of social media to their businesses during the COVID-19 pandemic. It consists of five (5) questions. The third portion, which comprises five (5) questions and uses dichotomous (Yes, No, and I don't know) alternatives to gather data on the level of SMEs' use of the social media in doing business during the lockdowns. The last section of the questionnaire used a 5-point Likert scale question that included strongly agreed (5 points), agree (4 points), neutral (3 points), disagree (2 points), and strongly disagreed (1 point) to gather information about the degree of impact of social media usage on SMEs businesses during the COVID-19 pandemic lockdown. To minimise a lengthy questionnaire, only a few questions directly related to the study's goals were asked.

In ensuring validity, a pilot study was done on ten (10) Small and Medium-sized Enterprise business owners and shareholders to ensure that the construct of the questionnaire instruments was understood. According to the findings of the pilot study, some people had difficulty grasping a couple of the devices. Appropriate changes were made to the parts of the questionnaire that these specimens struggled with. Thereafter, the questionnaire was distributed directly to the study's chosen sample and the questions on average took about 15 minutes to complete. Owners & managers of Small and Medium-sized Enterprise firms in Ikotun and Ikeja were given the questionnaires. Permission and agreement to participate in the distribution were obtained and granted prior to the distribution.

A total of 190 questionnaires were provided for the survey by two employed research advisers (one for each study region) who assisted in the administration of the questions. With regards to the study participants, any SMEs whose owners and workers were not available at the firms within the time allotted for distribution of the questionnaires were skipped. Within the study's time frame, a total of 183 analysable versions of the questionnaire were retrieved. Given that 190 copies of the questionnaire were distributed, and 183 copies of the questionnaire were retrieved, the completion rate, which is the rate of the retrieved and distributed replicas of the questionnaires, was determined to be 96.3%, which is a very good completion percentage for any study's survey.

Likewise, the internal consistency reliability was tested using Cronbach's alpha as a reliability indicator for this study. The purpose of a reliability test is to determine how acceptable the data's internal and external regularity is and Table 1 shows the reliability coefficients. If Cronbach's alpha () >=0.7, the internal consistency is deemed adequate; if the test applied is 0.8 or above, the external reliability is considered good. The Cronbach's alpha ($\alpha$) are 0.763 and 0.804 as measured by standardised items of this reliability test. Because they are equal and greater than 0.7 and 0.8 respectively, they could be regarded as great and acceptable, indicating that the questionnaire instruments are internally consistent and free of bias from respondents as well as researcher error.

**Table 1** Showing the SPSS output of the reliability statistics

| Reliability Statistics | | |
|---|---|---|
| Cronbach's Alpha | Cronbach's Alpha Based on Standardized Items | N of Items |
| .763 | .804 | 26 |

**3.2 Data analysis and results**
This section analyses of the study results were presented beginning with the presentation of socio-demographic information of respondents followed by that of research questions and discussions on the findings.

**3.2.1 Socio-demographic Information of the Respondents**
Table 2 represents the socio-demographic characteristics of respondents. From the analysis of gender, the male is 116 (62.4%) and female were 70 (37.6%). This is an indication that males are more than females in the study. From the analysis of age group, respondents less than 20 years were 20 (10.8%), 20-29 years were 40 (21.5%), 30-39 yeas were 30 (16.1%), 40-49 were 43 (23.1%), 50-59 years 40 (21.5%), and 60years and above were 13 (7.0%). This is an indication that majority of the respondents are between the ages of (40 and 49) years. From the analysis of educational levels, respondents with no formal education were 11 (5.9%), those with primary education were 11 (5.9%), secondary education were 66 (35.5%), and tertiary education were 98 (52.7%). This is an indication that those with tertiary education were greater in number.

**Table 2:** Sociodemographic characteristics

| S.no | Variables | Frequency (F) | Percentage (%) |
|---|---|---|---|

| | Total valid responses = 183 | | |
|---|---|---|---|
| 1. | **Gender** | | |
| | Male | 116 | 62.4 |
| | Female | 70 | 37.6 |
| 2. | **Age group** | | |
| | Less than 20years | 20 | 10.8 |
| | 20-29years | 40 | 21.5 |
| | 30-39years | 30 | 16.1 |
| | 40-49years | 43 | 23.1 |
| | 50-59years | 40 | 21.5 |
| | 60years above | 13 | 7.0 |
| 3. | **Level of Education** | | |
| | No formal Education | 11 | 5.9 |
| | Primary | 11 | 5.9 |
| | Secondary | 66 | 35.5 |
| | Tertiary | 98 | 52.7 |
| 4. | **Marital Status** | | |
| | Single | 91 | 48.9 |
| | Married | 46 | 24.7 |
| | Separated | 27 | 14.5 |
| | Widow(er). | 22 | 11.8 |
| 5. | **Religion** | | |
| | Christian | 72 | 38.7 |
| | Muslims | 88 | 47.3 |
| | None | 26 | 14.0 |

**Table 2:** Socio-demographic characteristics (Contd…)

| S.no | Variables | Frequency (F) | Percentage (%) |
|---|---|---|---|
| | Total valid responses =183 | | |
| 6. | **Workforce size** | | |
| | Less than 5 | 68 | 36.6 |
| | 5-50 | 101 | 54.3 |
| | 51-90 | 8 | 4.3 |
| | 91 to 250 but not more than. | 9 | 4.8 |
| 7. | **Category** | | |
| | Electronics stores | 69 | 37.1 |
| | Boutiques | 51 | 27.4 |
| | Shopping malls | 66 | 35.5 |
| 8. | **Location** | | |
| | Ikeja | 90 | 49.2 |
| | Ikotun | 93 | 50.8 |

From the analysis of the marital status, single respondents were 91 (48.9%), married respondents were 46 (24.7%), separated respondents were 27 (14.5%), and widow (er) were 22 (11.8%). This is an indication that majority of the respondents were single. From the analysis of religion, 72 (38.7%) respondents were Christians, 88 (47.3%) respondents were Muslims, and 26 (14.0%) respondents belong to no religion. This indicates that respondents who are Muslims are in the majority. The analysis of workforce size shows that workforce size that is less than 5 were 68 (36.6%), 5-50 were 101 (54.3%), 51-90 were 8 (4.3%) and 91-250 were 9 (4.8%). From the analysis of categories, electronic stores were 69 (37.1%), boutiques were 51 (27.4%), and shopping malls were 66 (35.5%). Indicating that electronic stores were the majority. The analysis of location showed that businesses in Ikeja were 90 (49.2%), and Ikotun 93 (50.8%). Indicating that Ikotun has the highest number of business clusters.

**3.2.2 Awareness of Nigeria SMEs on the usefulness of social media to their businesses during the pandemic**

Table 3 examined the awareness of Nigeria SMEs on the usefulness of social media on their businesses during the pandemic. In analysing whether respondents were aware they can do their business on social media before the pandemic? 115 (61.8%) said Yes, 55 (29.6%) said No, and 16 (8.6%) said they were Unaware. Indicating that most respondents were aware you can do your business on social media before the pandemic. In analysing whether respondents know there are some social media marketing features that facilitate sales of goods and services, 102 (54.8%) said Yes, 48 (25.8%) said No and 38 (19.3%) said they were Unaware. Indicating that most respondents are aware of some social media marketing features that facilitate sales of goods and services. In analysing whether some businesses thrived more during the pandemic lockdown due to the use of social media, 113 (60.8%) said Yes, 53 (28.5%) said No, and 20(11.7%) said they were Unaware. Indicating that most respondents were aware some businesses thrived more during the pandemic lockdown due to the use of social media. In analysing whether those respondents were aware that social media marketing platforms such as Facebook, Twitter, Instagram, Tik-Tok, and so on are useful for business branding and development, 115 (61.8%) said Yes, 37 (19.9%) said No and 34(18.5%) said they are Unaware. Indicating that most of those respondents are aware that social media marketing platforms such as Facebook, Twitter, Instagram, Tik-Tok, and so on are useful for business branding and development. In analysing whether respondents were aware they can do their business on social media before the COVID-19? 115 (61.8%) said Yes, 55 (29.6%) said No, and 16 (8.6%) said they are Unaware. Indicating that most respondents were aware you can do your business on social media before the pandemic. In analysing whether respondents were aware that the pandemic presented businesses with enormous opportunities on social media? 115 (61.8%) said Yes, 51 (27.4%) said No, and 20 (11.7%) said they are Unaware. Indicating that most respondents were aware that the pandemic presented businesses with enormous opportunities on social media.

**Table 3**: Responses to awareness of Nigeria SMEs on the usefulness of social media to their business during the pandemic (Total Respondents = 183)

| S.no | Questions and Options | Frequency (F) | Percentage (%) |
|---|---|---|---|
| 1. | **Were you aware you can do your business on social media before the covid-19?** | | |
| | Yes | 115 | 61.8 |
| | No | 55 | 29.6 |
| | I am not aware | 16 | 8.6 |
| 2. | **Do you know there are some social media marketing features that facilitate sales of goods and service?** | | |
| | Yes | 102 | 54.8 |
| | No | 48 | 25.8 |
| | I am not aware | 38 | 19.3 |
| 3. | **Are you aware that some businesses thrived more during the covid-19 pandemic lockdown due to use of social media?** | | |
| | Yes | 113 | 60.8 |
| | No | 53 | 28.5 |
| | I am not aware | 20 | 11.7 |
| 4. | **Are you aware that social media marketing platforms such as Facebook, twitter, Instagram, Tik-Tok, and so on are useful for your business branding and development most especially in this kind of social distancing era?** | | |
|   | Yes | 115 | 61.8 |
|   | No | 37 | 19.9 |
|   | I am not aware | 34 | 18.5 |
| 5. | **Are you aware that the pandemic presented businesses enormous opportunities on social media?** |   |   |
|   | Yes | 115 | 61.8 |
|   | No | 51 | 27.4 |
|   | I am not aware | 20 | 11.7 |

### 3.2.3 Level of usage of social media for doing business during the pandemic

Table 4 examined the level of usage of social media for doing business during the pandemic. In analysing what informs respondents' social media usage before the pandemic? 48 (25.8%) said Business/Trading, 74 (39.8) said Entertainment/News, and 64 (34.4%) kept up with friends and family. In analysing what informs respondents on social media usage during the pandemic, 29 (15.6%) said Business/Trading, 61 (32.8%) said Entertainment/News, said 96 (51.6%) keeping up with friends and family. This indicates that entertainment/news informs most respondents' social media usage before the pandemic. In analysing how often respondents use social media as leverage for adverting and promoting their businesses during the national lockdowns; 98 (52.7%) respondents said regularly, 56 (31.7%) respondents said sometimes, 29 (15.6%) respondents said not interested.

This indicates majority of respondent were regularly leveraging the use of social media for advertising and promoting their businesses during the national lockdowns. In analysing how the respondents rate their leveraging social media use for sharing their various business brands during the national lockdown; 96 (51.6%) which is the highest distribution said they it's 'high', 55 (29.6%) said it is rated 'medium' and the lowest distribution 35 (18.8%) said they would rate the usage as 'low'. This is an indication that their leveraging social media use for sharing their various business brands during the national lockdown was high. In analysing how respondents rate their leveraging social media use for interacting with their customers and potential customers during the pandemic lockdown; 129 (69.4%) which is the highest distribution said they it's 'high', 43 (23.1%) said medium and 14 (6.5%) said low. This is an indication that their leveraging social media use for interacting with their customers and potential customers during the pandemic lockdown

**Table 4:** Responses to how SMEs leveraged the use of social media for doing business during the pandemic lockdowns (Total Respondents = 183).

| S.no | Questions and Options | Frequency (F) | Percentage (%) |
|---|---|---|---|
| 1. | **What informs your social media usage before the COVID-19 lockdown?** |   |   |
|   | Business/Trading | 48 | 25.8 |
|   | Entertainment/News | 74 | 39.8 |
|   | Keeping up with friends and family | 64 | 34.4 |
| 2. | **What informs your social media usage during the COVID-19 lockdown?** |   |   |
|   | Business/Trading | 29 | 15.6 |
|   | Entertainment/News | 61 | 32.8 |
|   | Keeping up with friends and family | 96 | 51.6 |
| 3. | **How often do you use your social media for adverting and promoting your business during the COVID-19 lockdown?** |   |   |
|   | Regularly | 98 | 52.7 |
|   | Sometimes | 56 | 31.7 |
|   | Not interested | 29 | 15.6 |
| 4. | **How will you rate your leverage on social media use for sharing your business brand during the national lockdown?** | | |
| | High | 96 | 51.6 |
| | Medium | 55 | 29.6 |
| | Low | 35 | 18.8 |
| 5. | **How will you rate your leverage on social media use for interacting with your customers and potential customers during the pandemic lockdown?** | | |
| | High | 129 | 69.4 |
| | Medium | 43 | 23.1 |
| | Low | 14 | 6.5 |

### 3.2.4 Degree of impact of social media usage on businesses

Table 5 examined the degree of impact that the usage of social media has on the selected businesses. In analysing how the use of social media for businesses during COVID-19 improved respondents' businesses, 13 respondents (7.0%) Strongly disagreed, 19 (10.2%), disagreed 25 (13.4%), Neutral, 60 (32.3%) Agree, 69 (37.1%) strongly agreed. This indicates that most respondents strongly agreed that the use of social media for business during the pandemic improved their businesses. In analysing whether respondents' social media usage has effectively promoted and helped developed respondents' brands. 7 (3.8%) Strongly disagree, 8 (4.3%), disagreed, 27 (14.5%), were Neutral, 68 (36.6%) Agree, 76(40.9%) strongly agreed. This is an indication that most respondents strongly agreed that respondents' social media usage has effectively promoted and helped developed respondents' brands. In analysing whether the use of social media allows respondents to get first-hand customer feedback. 43 (23.1%) Strongly disagree, 40 (21.5%), 14 (7.5%), Neutral, 35 (18.8%) Agree, 54 (29.0%) strongly agreed.

**Table 5:** Responses to the degree of impact that the usage of this social media has on the selected businesses (Total Respondents = 183).

| S.no | Questions and Options | Frequency (F) | Percentage (%) |
|---|---|---|---|
| 1. | **The use of social media for my business during COVID-19 greatly improved my business.** | | |
| | Strongly disagree | 13 | 7.0 |
| | Disagree | 19 | 10.2 |
| | Neutral | 25 | 13.4 |
| | Agree | 60 | 32.3 |
| | Strongly agree | 69 | 37.1 |
| 2. | **Social Media usage has effectively promoted, and help developed my brand.** | | |
| | Strongly disagree | 7 | 3.8 |
| | Disagree | 8 | 4.3 |
| | Neutral | 27 | 14.5 |
| | Agree | 68 | 36.6 |
| | Strongly agree | 76 | 40.9 |
| 3. | **The use of social media allows me to get first-hand customer feedback.** | | |
| | Strongly disagree | 43 | 23.1 |
| | Disagree | 40 | 21.5 |
| | Neutral | 14 | 7.5 |
| | Agree | 35 | 18.8 |
| | Strongly agree | 54 | 29.0 |
| 4. | **The use of social media in doing business during COVID-19 further enhances my business connection and network.** | | |
| | Strongly disagree | 4 | 2.1 |
| | Disagree | 10 | 5.4 |
| | Neutral | 52 | 28.0 |
| | Agree | 51 | 27.4 |
| | Strongly agree | 69 | 37.1 |
| 5. | **The use of social media during COVID-19 created an opportunity to share feelings and opinions between me and my customers**. | | |
| | Strongly disagree | 15 | 8.1 |
| | Disagree | 9 | 4.8 |
| | Neutral | 21 | 11.3 |
| | Agree | 59 | 31.7 |
| | Strongly agree | 79 | 42.5 |

This is an indication that most respondents strongly agreed that the use of social media allows respondents to get first-hand customer feedback. In analysing whether the use of social media for business during the pandemic further enhanced business connection and network, 4 (2.1%) Strongly disagree, 10 (5.4%), 52(28.0%), Neutral, 51 (27.4%) Agree, 69 (37.1%) strongly agreed. This is an indication that most respondents strongly agreed that the use of social media for business during the pandemic further enhanced business connections and networks. In analysing whether the use of social media during the pandemic created an opportunity to share feelings and opinions between business owners and customers. 15 (8.1%) of respondents strongly disagreed, 9 (4.8%) disagreed, 21 (11.3%) were neutral, 59 (31.7%) agreed and 79 (42.5%) strongly agreed. This is an indication that most respondents strongly agreed that the use of social media during the pandemic created an opportunity to share feelings and opinions between business owners and customers.

### 3.3 Research question one

Table 6 presents the analyses of the awareness of Nigeria's SMEs on the usefulness of social media to their business during the COVID-19 pandemic. In examining whether respondents were aware they can conduct their businesses on social media before the COVID-19, a mean value of 1.4754 was observed which is greater than the median value of 1.0000 indicating that respondents are aware that the business can be done on social media before the pandemic.

In examining whether respondents know there are some social media marketing features that facilitate sales of goods and services, a mean value of 1.6230 was observed which is greater than 1.0000 indicating that respondents know there are social media features that facilitate sales of goods and services. In examining the awareness of respondents that some businesses thrived more during the COVID-19 pandemic lockdown due to the use of social media, a mean value of 1.4754 was observed which is greater than the median value of 1.000 indicating that respondents are aware. In examining respondents' awareness that social media marketing platforms are useful for business branding and development, especially in the social distancing era, a mean value of 1.5410 which is greater than the median value of 1.0000 was observed.

**Table 6** Descriptive statistical analysis of awareness of Nigeria SMEs on the usefulness of social media to their business during the COVID-19 pandemic

| | | Were you aware you can do your business on social media before the COVID-19? | Do you know there are some social media marketing features that facilitate sales of goods and service? | Are you aware that some businesses thrived more during the COVID-19 pandemic lockdown due to use of social media? | Are you aware that social media marketing platforms such as Facebook, twitter, Instagram, Tik-Tok etc. are useful for your business branding and development most especially in this kind of social distancing era? | Are you aware that the pandemic presented businesses enormous opportunities on social media? |
|---|---|---|---|---|---|---|
| N | Valid | 183 | 183 | 183 | 183 | 183 |
| Mean | | 1.4754 | 1.6230 | 1.4754 | 1.5410 | 1.4645 |
| Median | | 1.0000 | 1.0000 | 1.0000 | 1.0000 | 1.0000 |
| Std. Deviation | | .65314 | .77383 | .66150 | .76834 | .66100 |

**Source:** SPSS IBM version 20, Note: Median serves as the cut-off point. The decision will be made based on the Mean of the responses.

This indicates respondents are aware of the usefulness of social media for business branding. In analysing the awareness of respondents on the enormous opportunities social media presented. The mean value of 1.4645 which is greater than 1.0000 was observed indicating that respondents are aware of the business opportunities presented by social media. This generally indicates a high level of awareness of Nigerian SMEs on the usefulness of social media to their business during the COVID-19 pandemic.

From the analysis on Table 7 below, awareness of Nigeria SMEs on the usefulness of social media to their business during the COVID-19 pandemic, a mean value of 1.6699 was observed which is greater than the median value of 1.5000 therefore Nigerian SMEs are aware of the usefulness of social media to their business during the pandemic.

**Table 7** Summary of descriptive statistical analysis of awareness of Nigeria SMEs on the usefulness of social media to their business during the COVID-19 pandemic

| Statistics | | |
|---|---|---|
| AWARENESS | | |
| N | Valid | 183 |
| Mean | | 1.6699 |
| Median | | 1.5000 |
| Std. Deviation | | .45443 |

**Source:** SPSS IBM version 20, Note: Median serves as the cut-off point. Decision will be made based on the Mean of the responses.

### 3.4 Research question two
Table 8 analyses how SMEs leveraged the use of social media in doing business during the pandemic. In examining how often SMEs use social media for adverting and business promotion during the COVID-19 lockdown, a mean value of 1.62 was observed which is greater than the median value of 1.00. This indicates that respondents were frequently using social media as leverage for adverting and promoting their business during the COVID-19 lockdown. In examining respondents' rating of their ability to leverage social media use for sharing their business brand during the national lockdown, a mean value of 1.65 which is greater than the median value of 1.00 was observed. This indicates a high leverage of social media use for sharing their business brand during the national lockdown. In examining how respondents rate their leveraging social media use for interacting with your customers and potential customers during the pandemic national lockdown, a mean value of 1.36 which is greater than 1.00 indicates high leveraging social media for interacting with their customers and potential customers during the national lockdown. Generally, these analyses indicate that the SMEs leveraged greatly the use of social media in doing business during the pandemic lockdowns.

**Table 8** Summary of descriptive statistical analysis of the level of usage of social media in doing business during the COVID-19 pandemic

| | | How often do you use your social media for adverting and promoting your business during the covid-19 lockdown? | How will you rate your leverage on social media use for sharing your business brand during the national lockdown? | How will you rate your leverage on social media use for interacting with your customers and potential customers during the pandemic lockdown? |
|---|---|---|---|---|
| N | Valid | 183 | 183 | 183 |
| Mean | | 1.6230 | 1.6503 | 1.3552 |
| Median | | 1.0000 | 1.0000 | 1.0000 |
| Std. Deviation | | .74489 | .76178 | .59259 |

**Source:** SPSS IBM version 20, Note: Median serves as the cut-off point. Decision will be made based on the Mean of the responses.

Table 9 further analysed how SMEs leveraged the use of social media in doing business during the pandemic lockdowns. The Mean value derived was 4.3262 which is greater than the median value of 4.0000. Therefore, there is high leverage of social media in doing business by the SMEs during the pandemic lockdowns.

**Table 9** Summary of descriptive statistical analysis of how SME businesses leveraged on the use of social media for doing business during the pandemic National lockdown

| Statistics | | |
|---|---|---|
| USAGE | | |
| N | Valid | 183 |
| Mean | | 4.3262 |
| Median | | 4.0000 |
| Std. Deviation | | .86683 |

**Source:** SPSS IBM version 20, Note: Median serves as the cut-off point. The decision will be made based on the Mean of the responses. Questions 2, 3, and 5 were computed to form the summary.

Furthermore, Wilcoxon signed-rank test was used to compare the scores that come from the same participants. The test returns the mean difference, standard deviation, median (Table 11) as well as the Z statistics, which is requires for N>30.

Table 10 indicates that 30 participants experienced higher yield using their on-site method as opposed to social media tools introduced during COVID, 97 attest that social media tool's introduction to their business resulted in overall business improvement, while the remaining 59 are indifferent. The table output can be expressed as:

**Table 10:** Social media usage before COVID-19 pandemic lockdown

|  |  | N | Mean Rank | Sum of Ranks |
|---|---|---|---|---|
| ATSMBAC - 14. Social media usage before COVID-19 pandemic lockdown? | Negative Ranks | 30[a] | 49.65 | 1489.50 |
|  | Positive Ranks | 97[b] | 68.44 | 6638.50 |
|  | Ties | 59[c] |  |  |
|  | Total | 186 |  |  |

a. ATSMBAC < 14. Social media usage before COVID-19 pandemic lockdown?
b. ATSMBAC > 14. Social media usage before COVID-19 pandemic lockdown?
c. ATSMBAC = 14. Social media usage before COVID-19 pandemic lockdown?
Where: ATSMBAC = Average Total Social media tools in business after COVID-19 lockdown.

The result on table 10 indicates that the post social media tools introduction show an increased business performance (average rank of 49.65 vs. average rank of 68.44).

a. Post social media score < pre social media score
b. Post social media score > pre social media score
c. Post social media score = pre social media score

**Table 11:** Descriptive Statistics Social media usage before COVID-19 pandemic lockdown

|  | N | Mean | Std. Deviation | Minimum | Maximum | Percentiles | | |
|---|---|---|---|---|---|---|---|---|
|  |  |  |  |  |  | 25th | 50th (Median) | 75th |
| 14. Social media usage before COVID-19 pandemic lockdown? | 186 | 1.91 | .773 | 1 | 3 | 1.00 | 2.00 | 3.00 |
| ATSMBAC | 186 | 2.3737 | .58716 | 1.00 | 3.00 | 2.0000 | 2.5000 | 3.0000 |

Wilcoxon signed rank is used to report Z statistics. The Z statistics is used for sample size N>30.

**Table 12: Z Statistics**

|  |  |  | ATSMBAC - 14. Social media usage before COVID-19 pandemic lockdown? |
|---|---|---|---|
| Z |  |  | -6.337[b] |
| Asymp. Sig. (2-tailed) |  |  | .000 |
| Monte Carlo Sig. (2-tailed) | Sig. |  | .000 |
|  | 95% Confidence Interval | Lower Bound | .000 |
|  |  | Upper Bound | .016 |
| Monte Carlo Sig. (1-tailed) | Sig. |  | .000 |
|  | 95% Confidence Interval | Lower Bound | .000 |
|  |  | Upper Bound | .016 |

a. Wilcoxon Signed Ranks Test

b. Based on negative ranks.

c. Based on 186 sampled tables with starting seed 2000000.

The test statistics on Table 12 discloses the statistical significance between pre COVID-19 lockdown Social media tools in business adoption and post COVID-19 lockdown social media tools in business adoption. A Z value of -6.337 which result in p-value < 0.001 indicates a statistically significant difference (Z = -6.337, p < 0.001) in the business performance pre and post social media tool's introduction. The median outputs of 2.0 for pre social media and 2.5 for post social media tools introduction as indicated by the descriptive statistics table also confirms this difference. The null hypothesis that there is no difference between performance of SMEs use of social media tools pre and post COVID-19 pandemic lockdown is therefore rejected, while the alternative hypothesis that there exists statistically significant difference between the pre and post COVID-19 lockdown SMEs use of social media tools.

### 3.5 Research question three

From the analysis in Table 13 of the degree of impact that the usage of this social media has on the selected business, a mean value of 4.5262 was observed. This is greater than the median value of 4.0000 indicating a high degree of impact of social media on selected businesses.

**Table 13** Summary of descriptive statistical analysis of the degree of impact that the usage of this social media has on the selected business

| Statistics | | |
|---|---|---|
| Degree_of_Impact | | |
| N | Valid | 183 |
| | Missing | 3 |
| Mean | | 4.5262 |
| Median | | 4.0000 |
| Std. Deviation | | .86683 |

**Source:** SPSS IBM version 20, Note: Median serves as the cut-off point. Decision will be made based on the Mean of the responses.

## 4. Discussion of the findings

The analysis of the first research question revealed that Nigerian SMEs are aware of the usefulness of social media to their businesses during the COVID-19 pandemic. In line with our findings that revealed a high level of awareness of the usefulness of social media to their businesses during the COVID-19. Prior to this time, various SMEs in Nigeria focused on selling and marketing their products physically without any interest in social media and lacking the knowledge of the usefulness of social media for their businesses until the emergence of COVID-19 pandemic lockdown and social distancing measures. Businesses have discovered various uses of social media for businesses transactions. The various feature which can facilitate the use of social media is available for example, businesses can be branded using social media platform like Facebook. The ability of business owners to reach a wider range of customers gives them an edge over businesses transacted traditionally. Payment can be made online by indicating the business account number where payment can be made. The usefulness of social media especially during the COVID-19 pandemic has made many businesses come up with new strategies of doing business without having to gather a large crowd of the customer to purchase their products.

The investigation of the second research question found that there is high leveraging the use of social media in doing business by the SMEs during the pandemic lockdowns. The SMEs leveraged social media for sharing their brands, advertisement, and promotion, as well as interacting with their customers and potential customers during the lockdowns. These findings are in tandem with that of Ali Taha *et al.* (2021) who discovered a high level of social media usage in doing business. The findings also depict other positive benefits which businesses enjoy when they utilise social media in doing their business. Frequent exposure through social media gives businesses a name in the market as well as informs the business owners of the new trends that they must imbibe in their business if they must thrive.

In Nigeria today, many people utilise social media in one way or the other to reach their friends and family, and those who have a business, commodity, or service to sell take the opportunity to market them through social media. Since there is no limit to what can be marketed on social media, for ease of transactions delivery of goods have also been adopted by businesses, and hence the issue of distance is no longer a barrier as customers can get whatever product they ordered delivered to them. Based on the findings of this research, respondents who utilise social media in doing their business before COVID-19 were exposed to entrepreneurial opportunities via social media during the pandemic. Ali Taha *et al.* (2021) also noted the existence of statistically significant differences in the use of social media among businesses during the first wave of the COVID-19 pandemic. The analysis carried out on the results addressing the research questions found that there is a high degree of impact of social media on businesses. These findings agree with Margado (2020) who found that in Slovakia, there is a marked e-commerce turnover increase, online sales between March 2019 and 2020 increased by 44%. Some businesses in Nigeria were found to have utilised social media to their benefit amidst the COVID-19 pandemic.

Businesses in Nigeria have been impacted by social media in various ways, for example, orders and sales can go on 24 hours even when the physical store is closed. The public can view displayed items from the comfort of their homes and make decisions on which to go for. The impacts social media have on businesses are enormous. However, despite the positive impacts there are some negative impacts such as the potentials of customers sending virus using unverified links to the business owner. Therefore, there is a need for SMEs to be intentional about the usage of social media in doing their business to avoid any form of loss and malicious attack.

**5. Conclusion and recommendation**

Businesses in Nigeria are aware that social media is useful to their business during COVID-19 pandemic and utilised it. The SMEs leveraged social media for sharing their brands, advertising, and promotion, and interacting with their customers and potential customers during the lockdown thereby increased the volume of business transactions. A high-level awareness also helped business owners on the best ways to utilise social media to impact their businesses in various ways. There is therefore a need for business owners who transact their business traditionally to adopt the usage of social media.

Based on the findings of this study it is recommended that labour/trade unions should create awareness by conducting campaigns to inform business owners about the benefits of utilising social media in carrying out their business transactions. Educate business owners on the best ways to use social media to benefit their business as well as the best tools and applications which can be favourable in doing business. Government should also make policies that will encourage the use of social media in doing business like encouraging communication companies to reduce broadband costs.